\documentclass[onecolumn,showpacs]{revtex4}
\usepackage{epsfig}
\usepackage{graphicx}%
\usepackage{amsmath}

\makeatletter
\def\btt#1{\texttt{\@backslashchar#1}}%
\DeclareRobustCommand\bblash{\btt{\@backslashchar}}%
\makeatother

\begin{document}

\title{Casimir pistons with hybrid boundary conditions }% Force line breaks with \\
\
\author{Xiang-hua Zhai}

\author{Xin-zhou Li}\email{kychz@shnu.edu.cn}

\affiliation{Shanghai United Center for Astrophysics(SUCA), Shanghai
Normal University, 100 Guilin Road, Shanghai 200234, China
}%
\date{\today}% It is always \today, today, but you may specify any date with \date.

\begin{abstract}
 The Casimir effect giving rise to an attractive or repulsive force
 between the configuration boundaries that confine the massless scalar
 field is reexamined for one to three-dimensional pistons in this paper.
 Especially, we consider Casimir pistons
 with hybrid boundary conditions, where the boundary
 condition on the piston is Neumann and those on other surfaces are
 Dirichlet. We show that the Casimir force on the piston is always
 repulsive, in contrast with the same problem
where the boundary conditions are Dirichlet on all
 surfaces.

\end{abstract}

% PACS, the Physics and Astronomy Classification Scheme.

\maketitle

\vspace{0.4cm}\noindent\textbf{1. Introduction} \vspace{0.4cm}

Casimir predicted that an attractive force should act between two
plane-parallel uncharged perfectly conducting plates in
vacuum\cite{casimir}. The force is due to the disturbance of the
vacuum of the quantized electromagnetic field under the existence of
the boundary conditions. This effect has now been verified by
precise measurements\cite{lamoreaux} and has been applied to the
fabrication of microelectromechanical systems(MEMS)\cite{serry}.
Casimir energies and forces have been calculated theoretically in
various different configurations including stratified media,
rectangular cavities, wedge, sphere, cylinder, sphere(lens) above a
disk and other\cite{bordag}. The calculations indicated that the
Casimir energy may change its sign depending not only on the
boundary conditions but also on geometry and topology of the
configuration. About the dependence of the Casimir force on the
boundary conditions, we have known that the Casimir force is still
attractive for two parallel infinitely permeable plates(associated
to Neumann-Neumann boundary conditions), which is identical to
Casimir's original result. However, the Casimir force between a
perfectly conducting plate and an infinitely permeable one
(associated to Dirichlet-Neumann boundary conditions) is
repulsive\cite{boyer}. Due to this peculiar aspect, the Casimir
effect with hybrid boundary conditions(Dirichlet-Neumann) has been
considered recent years\cite{kenneth}. As for the influence of
geometry and topology of the configuration to the Casimir force, it
has been claimed that the Casimir energy inside rectangular cavities
can be either positive or negative depending on the ratio of the
sides\cite{mamayev}. But the security of the conclusion needs to
suspect at least for two reasons: First, the calculations ignore the
divergent term associated with the boundary. With the term thrown
out, the cut-off technique can get the same finite result as what is
obtained by zeta function regularization which renormalizes the
surface term to zero. But physically, such a term cannot be
eliminated by a renormalization of the parameters of the
theory\cite{graham}. Second, it does not take into account the
nontrivial contribution to the vacuum energy from the outside region
of the box. Actually, these problems exist in the calculation of
Casimir energy of any single body. In contrast, there is no such
problems for the case of two rigid bodies if one only interested in
the force between them.

Two years ago, a modification of the rectangle-"Casimir piston"-was
introduced to avoid the above two problems\cite{cavalcanti}. The
two-dimensional Casimir piston consists of a single rectangle
divided into two by a partition(the "piston"). The Casimir force on
the piston is a well-defined finite force because the position of
the piston is independent of the divergent terms in the internal
vacuum energy and the external region. For a scalar field obeying
Dirichlet boundary conditions on all surfaces, when the separation
between the piston and one end of the cavity approaches infinity,
the force on the piston is towards another end(the closed end), that
is, the force is attractive. Later, the attractive Casimir force on
the piston again obtained for a three-dimensional electromagnetic
field with the perfect-conductor conditon\cite{hertzberg} and for a
three-dimensional scalar field with Dirichlet boundary conditions on
all surfaces\cite{edery}. But attraction does not occur in all
Casimir piston scenarios. In a recent paper\cite{barton}, the
Casimir piston in a cylinder for a weakly reflecting dielectric was
considered and it was shown that the force could switch from
attraction to repulsion with the plate separation increasing. Again,
some examples of repulsive Casimir pistons have been simply
discussed in a recent preprint\cite{fulling}.

On the other hand, it may be worth emphasizing that the Hurwitz zeta
function $\zeta(\nu;s)$ is the direct zeta function associated with
the hybrid boundary conditions, while the Riemann zeta function
$\zeta(s)$ is the direct zeta function associated with the Dirichlet
boundary condition. $\zeta(\nu;s)$ is a generalization of $\zeta(s)$
defined by

\begin{equation}
\zeta(\nu;s)=\sum_{n=0}^{\infty}\frac{1}{(n+\nu)^s},\hspace{1cm}(0<\nu\leq
1,\textrm{Re}s>1)
\end{equation}

\noindent It is obvious that $\zeta(s)=\zeta(1;s)$. For
$\textrm{Re}s>0$, one has\cite{li}

\begin{equation}
\Gamma(s)=(n+\nu)^s\int_0^{\infty}x^{s-1}e^{-(n+\nu)x}dx
\end{equation}

\noindent Therefore,

\begin{equation}
\zeta(\nu;s)=\frac{1}{\Gamma(s)}\int_0^{\infty}\frac{x^{s-1}e^{-(\nu-1)x}}{e^x-1}dx
\end{equation}

\noindent if the inversion of the order of summation and integration
can be justified; and this is guaranteed by the absolute convergence
if $\textrm{Re}s>1$. Now one can consider the integral

\begin{equation}
\zeta(\nu;s)=\frac{-e^{i\pi s}\Gamma(1-s)}{2\pi
i}\int_C\frac{z^{s-1}e^{-\nu z}}{1-e^{-z}}dz
\end{equation}

\noindent where the contour $C$ starts at infinity on the positive
real axis, encircles the origin once in the positive direction
excluding the point $\pm2i\pi,\pm4i\pi,\cdot\cdot\cdot$, and returns
to positive infinity. We can take $C$ to consist of real axis from
$+\infty$ to $r$ ($0<r<2\pi$), the circle $\mid z\mid=r$, and the
real axis from $r$ to $+\infty$. On making $r\rightarrow 0$, we have
obtained Eq.(4). Eq.(4) provides the analytic continuation of
$\zeta(\nu;s)$ over the whole plane, and $\zeta(\nu;s)$ is regular
everywhere except for a simple pole at $s=1$ with residue $1$.
Expanding the loop to infinity, the residues are at $\pm 2mi\pi$;
hence, if $\textrm{Re}s<0$, we have

\begin{equation}
\zeta(\nu;s)=\frac{2\Gamma(1-s)}{(2\pi)^{1-s}}\Big[\textrm{sin}\frac{1}{2}\pi
s\sum_{m=1}^{\infty}\frac{\textrm{cos}2m\pi\nu}{m^{1-s}}+\textrm{cos}\frac{1}{2}\pi
s\sum_{m=1}^{\infty}\frac{\textrm{sin}2m\pi\nu}{m^{1-s}}\Big]
\end{equation}

In this paper, we consider Casimir pistons for a massless scalar
field with hybrid boundary conditions. That is, on the surface where
the piston lies, the boundary condition is Neumann, and the boundary
conditions are Dirichlet on other surfaces. We discuss one to three
dimensional pistons using generalized zeta function regularization
technique. Due to the existence of hybrid boundary conditions,
Hurwitz zeta functions and Epstein zeta functions emerge naturally.
In one dimensional case, it is very easy to get the analytic result
by regularizing Hurwitz zeta function directly. In two and three
dimensional cases, we need to do the calculation numerically after
the regularization. In all of the three cases, we show that the
Casimir force on the piston with hybrid boundary conditions are
repulsive. With the separation increasing, the force on the piston
decreases rapidly.

\vspace{0.4cm} \noindent\textbf{2. One-dimensional piston}
 \vspace{0.4cm}

Consider a quantized scalar field constrained in the interval $L$ on
the real line(See Fig.1). There is a point in the interval where the
piston lies. The piston divides the interval into two labeled $A$
and $B$. The distance between the piston and the left point is $a$.
The total energy of the vacuum for the system can be written as the
sum of three terms:

 \begin{equation}
 E=E^A(a)+E^B(L-a)+E^{out}
  \end{equation}

\noindent  where $E^A(a)$ and $E^B(L-a)$ are given by the results
through cut-off technique, which consist of divergent terms and
finite terms, where the finite terms are the same as what are
obtained by zeta function regularization denoted $E_R^A(a)$ and
$E_R^B(L-a)$. The divergent terms and the energy from the exterior
region in the total energy are independent of the position of the
piston\cite{cavalcanti}, so the Casimir force on the piston is as
follows:

\begin{equation}
F=- \frac{\partial}{\partial a}[E_R^A(a)+E_R^B(L-a)]
\end{equation}

\noindent With Dirichlet boundary condition on one point and Neumann
on the other point, the eigenfrequencies in interval $A$ are

\begin{equation}
\omega_n=\frac{\Big(n+\frac{1}{2}\Big)\pi}{a},\hspace{0.5cm}n=0,1,2,\cdot\cdot\cdot
\end{equation}

\noindent So the vacuum energy is given by($\hbar=c=1$ where $c$ is
the speed of light)

\begin{equation}
E(a)=\frac{1}{2}\sum_{n=0}^{\infty}(n+\frac{1}{2})\frac{\pi}{a}
\end{equation}

\noindent Using zeta function regularization, we start with the
function

\begin{equation}
\mathcal
{E}(a;s)\equiv\frac{\pi}{2}\sum_{n=0}^{\infty}\Big[\Big(n+\frac{1}{2}\Big)\frac{1}{a}\Big]^{-s}
\end{equation}

\noindent which is defined for $\textrm{Re}(s)>1$. We will see in
the following that its analytic continuation to the complex s-plane
is well-defined at $s=-1$. So we can write the regularized Casimir
energy as $E_R^A(a)=\mathcal {E}(a;-1)$.

Eq.(10) can be rewritten as

\begin{eqnarray}
\mathcal {E}(a;s)&=&\frac{\pi a^s}{2}\zeta(\frac{1}{2};s)
\end{eqnarray}

\noindent Using Mellin transformation one can find

\begin{equation}
\Gamma\Big(\frac{s}{2}\Big)\pi^{-s/2}(2^s-1)^{-1}\zeta(s)=\Gamma\Big(\frac{1-s}{2}\Big)\pi^{(s-1)/2}(2^{1-s}-1)^{-1}\zeta(1-s)
\end{equation}

\noindent Taking $s=-1$, we get

\begin{equation}
\zeta(\frac{1}{2};-1)=\frac{1}{12\pi^2}\zeta(\frac{1}{2};2)=\frac{1}{24}
\end{equation}

\noindent and so

\begin{equation}
E_R^A(a)=\mathcal {E}(a;-1)=\frac{\pi}{48a}
\end{equation}

\noindent The corresponding expression for interval $B$ is

\begin{equation}
E_R^B(L-a)=\frac{\pi}{48(L-a)}
\end{equation}

\noindent Thus, taking the limit $L\rightarrow\infty$, we obtain the
repulsive Casimir force on the piston

\begin{equation}
F=\frac{\pi}{48a^2}
\end{equation}

The result is the same as what was obtained in \cite{fulling} where
an exponential cutoff technique was used.

\vspace{0.4cm} \noindent\textbf{3. Two-dimensional piston}
 \vspace{0.4cm}

As illustrated in Fig. 2, the rectangle is divided into two by the
piston. The boundary condition on the piston is Neumann, and those
on other surfaces are Dirichlet. Similar to one-dimensional case,
the Casimir force acting on the piston is:

\begin{equation}
F=-\frac{\partial}{\partial a}\Big[E_R^A(a,b)+E_R^B(L-a,b)\Big]
\end{equation}

\noindent The vacuum energy of area $A$ is

\begin{equation}
E(a,b)=\frac{1}{2}\sum_{m=0}^{\infty}\sum_{n=1}^{\infty}\sqrt{\Big(m+\frac{1}{2}\Big)^2\Big(\frac{\pi}{a}\Big)^2+\Big(\frac{n\pi}{b}\Big)^2}
\end{equation}

\noindent In order to calculate the summation $E(a,b)$, we can
consider the more general expression $\mathcal {E}(a,b;s)$ as
follows

\begin{eqnarray}
\mathcal
{E}(a,b;s)&\equiv&\frac{\pi}{2}\sum_{m=0}^{\infty}\sum_{n=1}^{\infty}\Big[\Big(m+\frac{1}{2}\Big)^2\Big(\frac{1}{a}\Big)^2+\Big(\frac{n}{b}\Big)^2\Big]^{-s/2}\nonumber\\
&=&\frac{\pi}{8}\sum_{m,n=-\infty}^{\infty
\hspace{0.3cm}\prime}\bigg[\Big[\Big(\frac{m}{2a}\Big)^2+\Big(\frac{n}{b}\Big)^2\Big]^{-s/2}
-\Big[\Big(\frac{m}{a}\Big)^2+\Big(\frac{n}{b}\Big)^2\Big]^{-s/2}\bigg]-\frac{\pi a^s}{4}\zeta(\frac{1}{2};s)\nonumber\\
&=&\frac{\pi}{8}\Big[Z_2\Big(\frac{1}{2a},\frac{1}{b};s\Big)-Z_2\Big(\frac{1}{a},\frac{1}{b};s\Big)\Big]
-\frac{\pi a^s}{4}\zeta(\frac{1}{2};s)
\end{eqnarray}

\noindent Where $Z_p(a_1,\cdot\cdot\cdot,a_p;s)$ is Epstein zeta
function which is defined as
$Z_p(a_1,\cdot\cdot\cdot,a_p;s)\equiv\sum_{n_1,\cdot\cdot\cdot,n_p=-\infty}^{\infty
\hspace{0.4cm}\prime}\Big[(n_1a_1)^2+\cdot\cdot\cdot+(n_pa_p)^2\Big]^{-\frac{s}{2}}$
and the prime means that the term $n_1=n_2=\cdot\cdot\cdot=n_p=0$
has to be excluded. Applying the reflection formulae

\begin{equation}
(a_1\cdot\cdot\cdot
a_p)\Gamma\Big(\frac{s}{2}\Big)\pi^{-s/2}Z_p(a_1\cdot\cdot\cdot
a_p;s)=\Gamma\Big(\frac{p-s}{2}\Big)\pi^{(s-p)/2}Z_p(\frac{1}{a_1}\cdot\cdot\cdot
\frac{1}{a_p};p-s)
\end{equation}

\noindent and taking $s=-1$, we get

\begin{equation}
E_R^A(a,b)=\mathcal
{E}(a,b;-1)=-\frac{ab}{32\pi}\Big[2Z_2(2a,b;3)-Z_2(a,b;3)\Big]-\frac{\pi}{96a}
\end{equation}

\noindent When $a>b$, we can reexpress the Epstein zeta function
as\cite{ambjorn}

\begin{equation}
Z_2(a,b;3)=\frac{2\pi^2}{3a^2b}+\frac{16\pi}{ab^2}\sum_{m,n=1}^{\infty}\frac{n}{m}K_1\Big(2\pi
mn\frac{a}{b}\Big)+\frac{2\zeta(3)}{b^3}
\end{equation}

\noindent Where $K_n(z)$ is modified Bessel function . Substituting
Eq.(22) and the corresponding expression for $Z_2(2a,b;3)$ into Eq.
(21), we get

\begin{equation}
E_R^A(a,b)=-\frac{\zeta(3)a}{16\pi
b^2}-\frac{1}{2b}\sum_{m,n=1}^{\infty}\frac{n}{m}\Big[K_1\Big(4\pi
mn\frac{a}{b}\Big)-K_1\Big(2\pi mn\frac{a}{b}\Big)\Big]
\end{equation}

\noindent Inserting Eq. (23) and the corresponding expression for
$E_R^B(L-a,b)$ into Eq. (17) and taking $L\rightarrow\infty$, we
obtain the following result for the Casimir force on the piston

\begin{equation}
\lim_{L\rightarrow\infty}F=\frac{\pi}{b^2}\sum_{m,n=1}^{\infty}n^2\Big[2K_1^{\prime}\Big(4\pi
mn\frac{a}{b}\Big)-K_1^{\prime}\Big(2\pi mn\frac{a}{b}\Big)\Big]
\end{equation}

\noindent Where $K_1^{\prime}(z)=dK_1(z)/dz$. The numerical
calculation tells us that the force is positive and decreases
rapidly with the ratio $a/b$ increasing.

In the case that $a<b$, Eq. (22) changes to

\begin{equation}
Z_2(a,b;3)=\frac{2\pi^2}{3b^2a}+\frac{16\pi}{ba^2}\sum_{m,n=1}^{\infty}\frac{n}{m}K_1\Big(2\pi
mn\frac{b}{a}\Big)+\frac{2\zeta(3)}{a^3}
\end{equation}

 \noindent and the resulted force on the piston is

 \begin{equation}
 \lim_{L\rightarrow\infty}F=\frac{3b\zeta(3)}{32\pi a^3}-\frac{\pi}{96a^2}-\frac{\zeta(3)}{16\pi
 b^2}+\frac{\pi b}{4a^3}\sum_{m,n=1}^{\infty}n^2\Big[K_0\Big(\pi
 mn\frac{b}{a}\Big)-4K_0\Big(2\pi mn\frac{b}{a}\Big)\Big]
 \end{equation}

 \noindent It can also be shown that the force is repulsive and
 decreases with the value $a/b$ increasing. Furthermore, one can
 know from the numerical computation that Eq. (24) and Eq. (26) are
 connected that they have the same value of the force when $a=b$.

\vspace{0.4cm} \noindent\textbf{4. Three-dimensional piston}
 \vspace{0.4cm}

 Similarly, the results of two-dimensional pistons can be extended to those of three-dimensional pistons.
 The three-dimensional piston is depicted in Fig. 3, where again the
 boundary condition on the piston is Neumann and those on other
 surfacee are Dirichlet. For simplicity, we take the base as a square. The vacuum energy in cavity $A$ is

\begin{equation}
E(a,b,b)=\frac{1}{2}\sum_{m=0}^{\infty}\sum_{n_1,n_2=1}^{\infty}
\sqrt{\Big(m+\frac{1}{2}\Big)^2\Big(\frac{\pi}{a}\Big)^2+\Big(\frac{n_1\pi}{b}\Big)^2+\Big(\frac{n_2\pi}{b}\Big)^2}
\end{equation}

When $a>b$, we get the regularized vacuum energy in cavity $A$ as

\begin{equation}
E_R^A(a,b,b)=-\frac{a\beta(2)}{48b^2}+\frac{\zeta(3)a}{16\pi b^2}
+\frac{1}{2b}\sum_{m,n_1,n_2=1}^{\infty}\frac{\sqrt{n_1^2+n_2^2}}{m}\Big[K_1\Big(2\pi
m\sqrt{n_1^2+n_2^2}\frac{a}{b}\Big)-K_1\Big(4\pi
m\sqrt{n_1^2+n_2^2}\frac{a}{b}\Big)\Big]
\end{equation}

\noindent where $\beta(2)$ is a Dirichlet series defined as
$\beta(s)\equiv\sum_{n=0}^{\infty}(-1)^n(2n+1)^{-s}$ which comes
from the relation $Z_2(1,1;s)=4\zeta(s)\beta(s)$\cite{zucker} during
the regularization. Substituting Eq.(28) and the corresponding
expression for the regularized vacuum energy in cavity $B$ into the
following expression for Casimir force on the piston

\begin{equation}
F=-\frac{\partial}{\partial a}\Big[E_R^A(a,b,b)+E_R^B(L-a,b,b)\Big]
\end{equation}

\noindent and taking $L\rightarrow\infty$, we obtain the force on
the piston as

\begin{eqnarray}
\lim_{L\rightarrow\infty}F=\frac{\pi}{b^2}\sum_{m,n_1,n_2=1}^{\infty}(n_1^2+n_2^2)\Big[2K_1^{\prime}\Big(4\pi
m\sqrt{n_1^2+n_2^2}\frac{a}{b}\Big)-K_1^{\prime}\Big(2\pi
m\sqrt{n_1^2+n_2^2}\frac{a}{b}\Big)\Big]
\end{eqnarray}

The force is positive from the result of numerical calculation and
it approaches zero with the ratio of $a/b$ approaching infinity.

In the case that $a<b$, the regularized vacuum energy in cavity $A$
can be reexpressed as

\begin{eqnarray}
E_R^A(a,b,b)&=&\frac{7\pi^2b^2}{11720a^3}-\frac{3\zeta(3)b}{64\pi
a^2}+\frac{\pi}{192a}+\frac{1}{4a}\sum_{m,n=1}^{\infty}\frac{m}{n}\Big[K_1(\pi
m n\frac{b}{a})-2K_1(2\pi m n\frac{b}{a})\Big]\nonumber\\
&+&\frac{b^{1/2}}{2a^{3/2}}\sum_{m,n=1}^{\infty}\Big(\frac{m}{n}\Big)^{3/2}\Big[K_{3/2}(2\pi
m n\frac{b}{a})-\frac{\sqrt{2}}{4}K_{3/2}(2\pi m
n\frac{b}{a})\Big]\nonumber\\
&+&\frac{b^{1/2}}{2a^{3/2}}\sum_{m,n_1,n_2=1}^{\infty}\Big(\frac{m}{\sqrt{n_1^2+n_2^2}}\Big)^{3/2}\Big[K_{3/2}(2\pi
m \sqrt{n_1^2+n_2^2}\frac{b}{a})-\frac{\sqrt{2}}{4}K_{3/2}(2\pi m
\sqrt{n_1^2+n_2^2}\frac{b}{a})\Big]
\end{eqnarray}

\noindent Then the force on the piston is

\begin{eqnarray}
\lim_{L\rightarrow\infty}F&=&\frac{7\pi^2b^2}{3840a^4}-\frac{3\zeta(3)b}{32\pi
a^3}+\frac{\pi}{192a^2}-\frac{\beta(2)}{48b^2}+\frac{\zeta(3)}{16\pi
b^2}\nonumber\\
&-&\frac{\pi b}{4a^3}\sum_{m,n=1}^{\infty}m^2\Big[K_0(\pi m
n\frac{b}{a})-4K_0(2\pi m n\frac{b}{a})\Big]-\frac{\pi
b^{3/2}}{a^{7/2}}\sum_{m,n=1}^{\infty}\frac{m^{5/2}}{n^{1/2}}\Big[K_{1/2}(2\pi
m n\frac{b}{a})-\frac{\sqrt{2}}{8}K_{1/2}(\pi m
n\frac{b}{a})\Big]\nonumber\\
&-&\frac{\pi
b^{3/2}}{a^{7/2}}\sum_{m,n_1,n_2=1}^{\infty}\frac{m^{5/2}}{(n_1^2+n_2^2)^{1/4}}\Big[K_{1/2}(2\pi
m \sqrt{n_1^2+n_2^2}\frac{b}{a})-\frac{\sqrt{2}}{8}K_{1/2}(\pi m
\sqrt{n_1^2+n_2^2}\frac{b}{a})\Big]
\end{eqnarray}

\noindent The force is again repulsive and decreases with the ratio
$a/b$ increasing (see Fig.4).

For the special case that $a=b$, which means cavity $A$ is a cube,
we find from both Eq. (30) and Eq. (32) that the force on the piston
is (in unit $\hbar c$) $F=\frac{0.00041244}{b^2}$.

\vspace{0.4cm} \noindent\textbf{5. Conclusion}
 \vspace{0.4cm}

We discuss one to three-dimensional Casimir pistons for a massless
scalar field with hybrid boundary conditions, where the boundary
condition on the piston is Neumann and those on other surfaces are
Dirichlet. We find the forces on the pistons are always repulsive,
in contrast with the same problem where the boundary conditions are
Dirichlet on all surfaces.

The problem of hybrid boundary conditions we study here is in
analogue with the problem in electromagnetic field that the piston
is an infinitely permeable plate and the other sides of the cavity
are perfectly conducting ones or the opposite case that the piston
is a perfectly conducting plate and the other sides are infinitely
permeable ones. This problem may be connected with the study of
dynamical Casimir effect and may be applied to the fabrication of
MEMS, which needs further investigation.

\begin{figure}
\epsfig{file=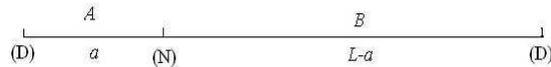,height=2.5in,width=3in}\caption{Casimir piston
in one dimensions.}
\end{figure}

\begin{figure}
\epsfig{file=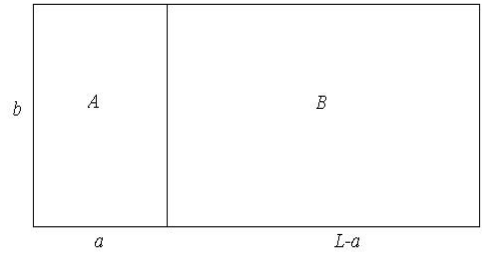,height=2.5in,width=3in} \caption{Casimir
piston in two dimensions.}
\end{figure}

\begin{figure}
\epsfig{file=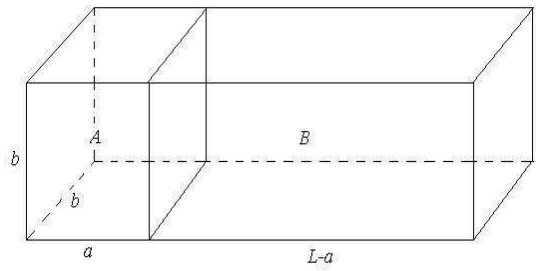,height=2.5in,width=3in} \caption{Casimir
piston in three dimensions.}
\end{figure}

\begin{figure}
\epsfig{file=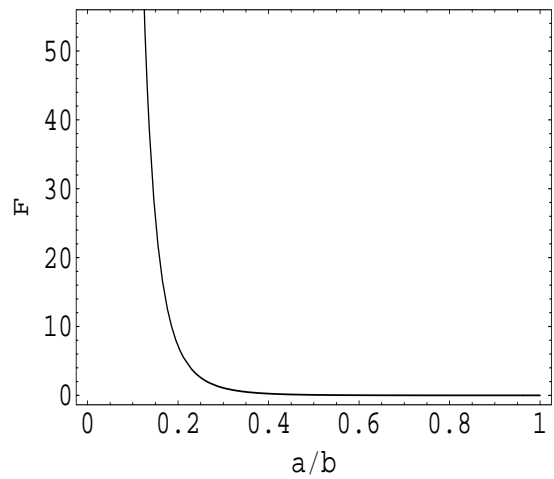,height=2.5in,width=3in} \caption{Casimir force
$F$ (in units $\hbar c/b^2$) on a three-dimensional piston versus
$a/b$ where $a$ is the plate separation and $b$ is the length of the
 sides of the square base. }
\end{figure}

\vspace{0.8cm} \noindent ACKNOWLEDGEMENT: This work is supported by
National Nature Science Foundation of China under Grant No. 10671128
and Shanghai Municipal Education Commission(No 06DZ005).

\end{document}